\documentclass[conference]{IEEEtran}
\usepackage{geometry}

\IEEEoverridecommandlockouts
\usepackage{cite}
\usepackage{amsmath,amssymb,amsfonts}
\usepackage{algorithmic}
\usepackage{graphicx}
\usepackage{textcomp}
\usepackage{xcolor}
\def\BibTeX{{\rm B\kern-.05em{\sc i\kern-.025em b}\kern-.08em
    T\kern-.1667em\lower.7ex\hbox{E}\kern-.125emX}}

\begin{document}
\newgeometry{top=72pt, bottom=54pt, left=54pt, right=54pt}

\title{Cat Royale: An Artistic Inquiry into Trust in Robots \\
}

\author{\IEEEauthorblockN{Matt Adams}
\IEEEauthorblockA{
\textit{Blast Theory}\\
Portslade, UK \\
matt@blasttheory.co.uk
}
\and
\IEEEauthorblockN{Nick Tandavanitj}
\IEEEauthorblockA{
\textit{Blast Theory}\\
Portslade, UK \\
nick.tandavanitj@blasttheory.co.uk}
\and
\IEEEauthorblockN{Steve Benford}
\IEEEauthorblockA{\textit{Mixed Reality Laboratory} \\
\textit{University of Nottingham}\\
Nottingham, UK \\
steve.benford@nottingham.ac.uk}
\and
\IEEEauthorblockN{Ayse Kucukyilmaz}
\IEEEauthorblockA{\textit{Mixed Reality Laboratory} \\
\textit{University of Nottingham}\\
Nottingham, UK \\
ayse.kucukyilmaz@nottingham.ac.uk}
\and
\IEEEauthorblockN{Victor Ngo}
\IEEEauthorblockA{\textit{Mixed Reality Laboratory} \\
\textit{University of Nottingham}\\
Nottingham, UK \\
victor.ngo@nottingham.ac.uk}
\and
\IEEEauthorblockN{Simon Castle-Green}
\IEEEauthorblockA{\textit{Mixed Reality Laboratory} \\
\textit{University of Nottingham}\\
Nottingham, UK \\
simon.castle-green@nottingham.ac.uk}
\and
\IEEEauthorblockN{Guido Salimberi}
\IEEEauthorblockA{\textit{Mixed Reality Laboratory} \\
\textit{University of Nottingham}\\
Nottingham, UK \\
guido.salimberi@nottingham.ac.uk}
\and
\IEEEauthorblockN{Pepita Bernard}
\IEEEauthorblockA{\textit{Mixed Reality Laboratory} \\
\textit{University of Nottingham}\\
Nottingham, UK \\
pepita.barnard@nottingham.ac.uk}
\and
\IEEEauthorblockN{Joel Fischer}
\IEEEauthorblockA{\textit{Mixed Reality Laboratory} \\
\textit{University of Nottingham}\\
Nottingham, UK \\
joel.fischer@nottingham.ac.uk}
\and
\IEEEauthorblockN{Alan Chamberlain}
\IEEEauthorblockA{\textit{Mixed Reality Laboratory} \\
\textit{University of Nottingham}\\
Nottingham, UK \\
alan.chamberlain@nottingham.ac.uk}
\and
\IEEEauthorblockN{Eike Schneiders}
\IEEEauthorblockA{\textit{University of Southampton}\\
Southampton, UK \\
eike.schneiders@soton.ac.uk}
\and
\IEEEauthorblockN{Clara Mancini}
\IEEEauthorblockA{
\textit{The Open University}\\
Milton Keynes, UK \\
clara.mancini@open.ac.uk}
}

\maketitle

\begin{abstract}
Cat Royale is an artwork created by the artists Blast Theory to explore the question of whether we should trust robots to care for our loved ones. The artists endeavoured to create a ‘Cat Utopia’, a luxurious environment that was inhabited by a family of three cats for six hours a day for twelve days, at the centre of which a robot arm played with them by wielding toys. Behind the scenes, the decision engine recommended games based on ongoing assessment of their happiness. A video installation featuring an eight-hour movie of the cats’ exploits is currently touring worldwide, provoking audiences to engage with the question of trust in autonomous systems.
\end{abstract}

\begin{IEEEkeywords}
Robots, cats, play, Cat Royale, art, animal-computer interaction
\end{IEEEkeywords}

\section{Introduction}

\begin{figure*}[t]
    \centering
    \includegraphics[width=\linewidth]{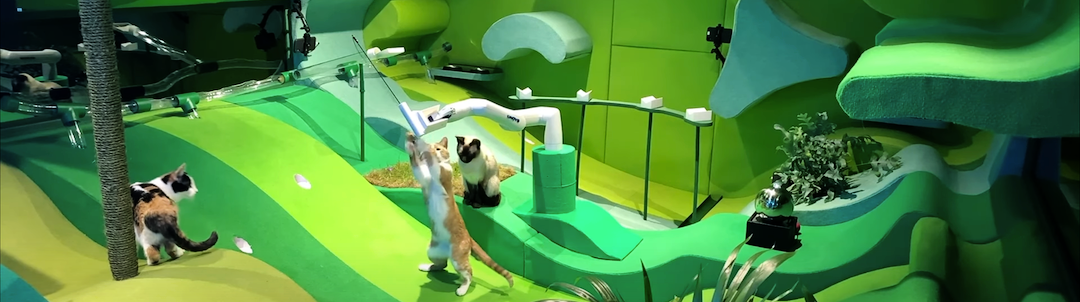}
    \caption{Inside a purpose-built enclosure, a robot arm lifts toys from four (magnetic) racks, wielding them to play with three cats, Clover, Pumpkin and Ghostbuster, who lived there for twelve days.}
    \label{fig:EnvDesign}
\end{figure*}

In 2022, the artists Blast Theory were engaged by the UKRI Trustworthy Autonomous Systems (TAS) Hub to create an artwork that would engage public audiences with the matter of trust in robots. Their response was Cat Royale, a unique artwork in which a robot attempts to play with a family of three cats inside a purpose-built enclosure over the course of twelve days in order to enrich their lives (see Figure-\ref{fig:EnvDesign}) . The cats' activities, including their 500 games with the robot, were filmed and streamed to visitors at the World Science Festival in Brisbane, Australia. The artists subsequently edited an eight-hour long movie which is now touring worldwide, while selected highlights were released online.

Should we trust robots to care for our loved ones?  Some might argue that care robots can fill gaps in strained social care systems. Others might worry about the replacement of human contact, job losses or safety. One area where such autonomous systems are already entering our lives is in caring for our companion animals. From automated feeders, to cat flaps, to toys, animals are increasingly being looked after by ‘robots’. Our animal companions also routinely encounter other robots in their environments that are not dedicated to looking after them, for example vacuum cleaners, lawnmowers, or food delivery robots. Given that many people care deeply about their companion animals, and that videos of animals engaging robots (e.g., cats riding vacuum cleaners) are popular online, the notion of deploying a robot to enrich the lives of companion animals, specifically cats, emerges as a compelling way of engaging the public with the question of trust, as well as posing a difficult technical challenge for roboticists to explore.

\section{Our presentation}

Our presentation will take the form of a video installation where conference delegates can watch edited highlights of the cats interacting with the robot, interspersed with reflections from the artists. Physically, this could range from a single large screen in the style of a digital poster, to a display in the background of a larger social area with relaxed seating for a slower audience experience.

The accompanying video briefly presents three example excepts of this material. The opening section introduces the project setting, its aims and format, and introduces the cats, the environment and robot arm. The middle section focuses on one important moment in Cat Royale in which the cat Clover physically engaged the robot in extended and vigorous play, ultimately ripping a toy away from its grip. The final section gives an example of artistic reflection on the work.

In the remainder of this accompanying paper we provide further background to Cat Royale, namely: an overview of its realisation; its touring history to date; and a summary of research findings that have emerged so far.

\section{An overview of Cat Royale}

Cat Royale comprised an enclosure, robot, orchestration system, a video capture system, and of course, the cats.

\subsection{The Enclosure} Cat Royale was housed in a bespoke enclosure,  designed to be a luxurious space that could cater to the cats' various needs, including providing ample sleeping dens, feeding areas, litter trays, a scratching post, cat grass and high perches and walkways. Its striking visual design was intended to convey the idea of luxury to the audience but was also decorated to help the individual cats stand out against the background for a computer vision system that tracked their movements.

\subsection{The Robot} Placed in the centre of the enclosure was the robotic arm. We chose a Kinova Gen3 lite robot arm, a small cobot with 6 degrees of freedom, with a 0.76 meters reach and an integrated two-finger gripper providing it with a maximum payload of only 0.5 kg. The short range and low payload were vital in order to ensure the safety of the cats. Through custom end-effectors and a wide range of different toys and attachments, the robot was able to offer a variety of games and treats to the cats. Every few minutes the robot's decision engine proposed a new game aimed at maximising a chosen cat's  happiness. The decision engine followed an exploration-to-exploitation cycle, starting with high exploration—incorporating random activities early on—before gradually shifting toward greater exploitation of the most popular activities later in the installation. If approved by the lead artist, the robot would pick up a toy and manipulate it in a way designed to engage the cats, while being monitored by the robot operator to ensure safety. To ensure animal welfare, each cat's individual stress score~\cite{kessler:turner:1997} was ranked at 15-minute intervals by the specialist in cat behaviour.

The robot offered more than 50 types of game, from simple props such as a cardboard box or a string, to wielding more elaborate toys such as an orange bird toy or a battery powered wiggling fish toy, to picking up balls and dropping them down a chute so they would roll across the floor. Following each game, the artists ranked each cat's engagement using the Participation in Play scale~\cite{Ellis:2022:Pip}, feeding this data back into the decision engine to refine its recommendations.\\

\subsection{The Cats} It was vitally important to acknowledge the needs of the three cats during the design of Cat Royale. Their welfare was our priority throughout the project. To ensure this, we recruited animal-computer interaction specialists, feline behaviourists, and veterinarians onto our team. A key cat recruitment criteria was the cats prior familiarity with each other, leading to higher comfort when sharing the enclosure. Furthermore, to ensure that the cats were not deprived of interaction with their owner, the owner moved to the art studio for 17 days (five cat habituation days and 12 days of deployment).


\subsection{The Orchestration System} Located behind a one-way glass mirror the artists, cat welfare specialist and robot operator monitored events, and intervened when necessary. The robot operator was responsible for manually triggering the robot to begin its next game at a safe and opportune moment and, at times, taking over manual control of the robot when unexpected events occurred such as toys becoming tangled up or the cats playing tug of war with them (and so stressing the arm) as we see in our video. Furthermore, all robot movement was dependent on the continuous pressing of a dead man's switch by the robot operator. In short, human improvisation was required to cope with the unpredictability of the cats and to ensure their safety as well as the safety and reliability of the robot.

\subsection{Video capture} All cat and robot activities within the enclosure were constantly filmed using eight cameras, in addition to the depth camera used for animal localisation and activity recognition. The output of these eight cameras were mixed in real-time by an experienced television vision mixer to capture the best possible view of the action and to add descriptive captions. This initial footage, comprising of 576 hours of video footage (12 days $\times$ 8 cameras $\times$ 6 hours/day), was subsequently edited into an eight-hour long film. Furthermore, a daily highlight video (between 3--5 minutes long) was edited for each day and presented to audiences globally using social media.

\section{Touring Cat Royale}

Cat Royale is significant as an artwork. At the time of writing, the video exhibition has appeared at.
\begin{itemize}
    \item The Curiocity Festival as part of the 2023 the World Science Fair in Brisbane, Australia (March-April 2023). 
    \item The Science Gallery, London, UK (June 2023-Jan 2024).
    \item The Wales Milennium Centre, UK, (January-Feb 2024)
    \item The Museum of Contemporary Art (MoCA), Taipei, as part of the Hello Human Exhibition (January-May 2024)
    \item The Teatr Studio, Warsaw, Poland, at the Palace of Culture as part of an exhibition about AI (2024).
\end{itemize}

This combination of touring installation and  social media highlights has led Cat Royale to be experienced by hundreds of thousands of people to date. The work also won the 2024 Webby Award in the category \textit{`People's Voice: AI, Metaverse \& Virtual Best Integrated Experience'}\footnote{https://winners.webbyawards.com/2024/ai-metaverse-virtual/ai-apps-and-experiences-features/best-integrated-experience/288063/cat-royale}. A systematic analysis of audience response is ongoing.

\section{Research from Cat Royale}

Cat Royale is also significant as a research project, and so should be of particular interest to conference delegates. Our research followed the method of ‘Performance-Led Research in the Wild’ \cite{Benford:ArtistLed:2013}, an open-ended, exploratory method in which creative artistic practice leads research, so that findings (and even research questions) eventually emerge from reflection on practice (contrasting with  hypothesis- or problem-led approaches from science and engineering). The following publications and findings have emerged from the work so far\footnote{Please note that space constraints preclude a review of related work, but that such reviews can be found in the following papers.}.

An initial overview of the work was published at the first conference on Trustworthy Autonomous Systems (TAS’23)~\cite{Schneiders:CatRoyale_TAS:2023}. A post-hoc reflection on how the artists designed and operated the experience published at ACM Computer-Human Interaction (CHI 2024) (best paper winner) raised the idea of designing 'multispecies robot worlds' \cite{Schneiders:2024:CatRoyaleRobotWorlds}. In addition to designing a robot's hardware and software, it is also necessary to design the wider world in which it will operate, for example robot-friendly accessories and interiors for our homes and workplaces. Moreover, these worlds need to be designed with the needs of animals (as well as people and robots) in mind, for example providing safe spaces from which to observe and choose to approach robots, or not.

A separate account of how the project negotiated an eighteen month ethical review process, also published at ACM CHI 2024 (best paper honorable mention), revealed key tensions in the ethics of multispecies technology research, also published at CHI 2024 (honourable mention award, top 5\%)~\cite{Benford:2024:CatRoyaleEthics}.

Cat Royale featured as a case study in a paper reflecting on how artistic practice and thinking can provoke robotics research with unusual challenges and ideas and lead to improvisation of new techniques, published at the 16th International Conference on Social Robotics (ICSR 2024)~\cite{benford2024artists}.

The first ever alt.HRI Video presentation at the 19th edition of the IEEE/ACM International Conference on Human-Robot Interaction (ACM HRI'24)~\cite{Schneiders:2024:altHRI} further highlighted the importance of multidisciplinarity in robotics research.

\bibliographystyle{ieeetr}
\bibliography{biblio}

\end{document}